\documentclass[12pt]{article} 

\usepackage{graphicx}
\title{Information Scientific and Educational Resource ``Electrophysics''}

\author{
\thanks{E-mail:GPAveryanov@mephi.ru}
G.P. Averyanov, V.V. Dmitrieva, N.P. Kornev, A.A. Fadeev\\\\
Department of Electrophysical Facilities, \\ National Research Nuclear University MEPhI, \\
Kashirskoye shosse 31, Moscow, 115409, Russia}
\date{ }
\begin{document} 
\maketitle
\newcounter{graf}
\begin{abstract}
The paper discusses an advanced level information system to support educational, research and scientific activities of the Department ``Electrophysical Facilities'' (DEF) of the National Research Nuclear University ``MEPhI'' (NRNU MEPhI), which is used for training of specialists of the course ``Physics of Charged Particle Beams and Accelerator Technology''. \end{abstract}
\section{Introduction}
In the development of modern IT, one of the most important and urgent tasks is the development of specialized information resources to support science and education in various subject areas, taking into account their specific features. The specifics of a specific subject area requires a serious adaptation of standard computer technologies commonly used in the scientific and educational departments of universities to the specific conditions and challenges.

After years of implementation of IT in educational and scientific activity at the NRNU MEPhI DEF, an open modular informational scientific-educational resource (INOR) “Electrophysics'' has been developed, intended for information support in solving scientific-practical tasks and training of specialists in the field of physics of charged particle beams and accelerator technology.

The INOR “Electrophysics'' supports several levels of access (Fig. 1):

\begin{itemize}
\item Standalone (or local) access from LAN computer classes of the DEF;
\item Corporate access from the classroom, campus and corporate networks of NRNU MEPhI;
\item Global access through Internet.
\end{itemize}

The local computer network of the Computing Laboratory of the DEF includes a computer classroom and a server with a network switch and a gateway to the Internet. Windows and GNU/Linux are used as operating systems. The Apache Web Server supports the operation of information website (www.accel.ru) and e-teaching portal (edu.accel.ru).

\begin{figure}[ht]
\begin{center}
\includegraphics[width=14.cm]{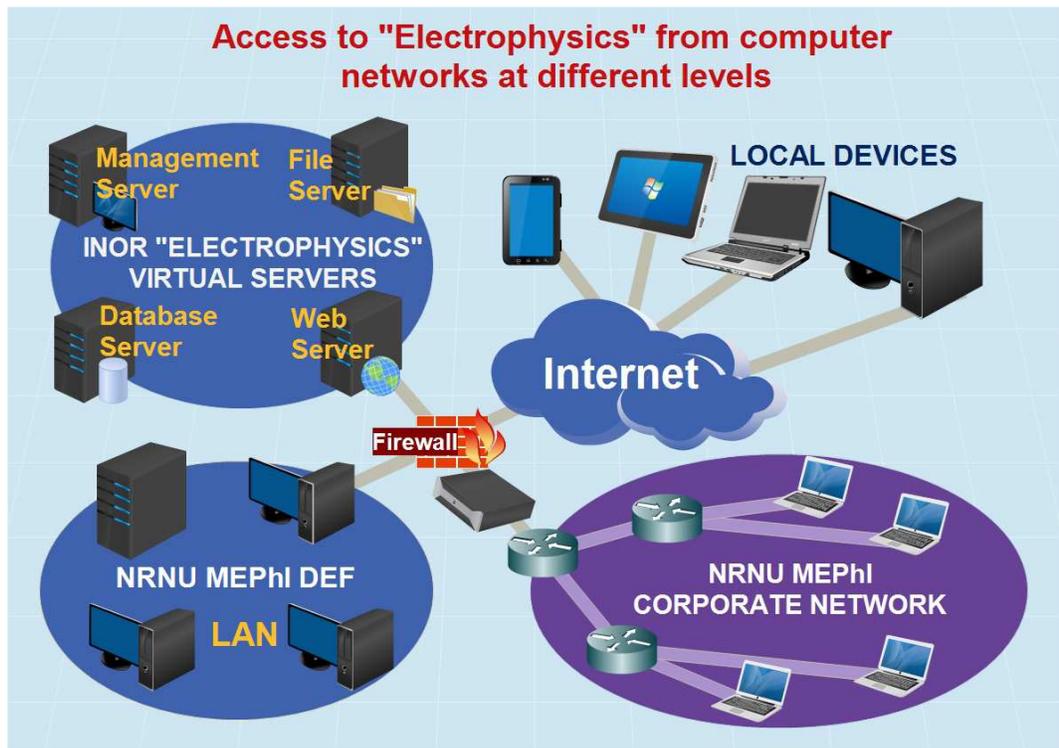}  
\end{center}
\caption{Levels of access to INOR ``Electrophysics''.}
\end{figure}

An informational website provides a remote access to the thematic content and specialized cross-platform web applications - simulation subsystems of the DEF used during the laboratory workshops.

E-teaching portal was created on the basis of the Virtual Teaching System Moodle. Protection of information resources from unauthorized access is provided through access control separately for students, teachers, and researchers.

\section{The concept of INOR ``Electrophysics''}

The concept of INOR ``Electrophysics'' is based on the principles and technologies of e-teaching and open education with the aim of providing information support of educational and scientific activities of the DEF in the training of specialists in the field of physics of charged particle beams and accelerator technology, and it can be considered in two aspects:
\begin{enumerate}
\item Object-oriented hierarchical (associated with hierarchy of thematic structuring of information resources);
\item Functional (associated with the information processes in applications implementing functions of the information system).
\end{enumerate}

\begin{figure}[ht]
\begin{center}
\includegraphics[width=16.cm]{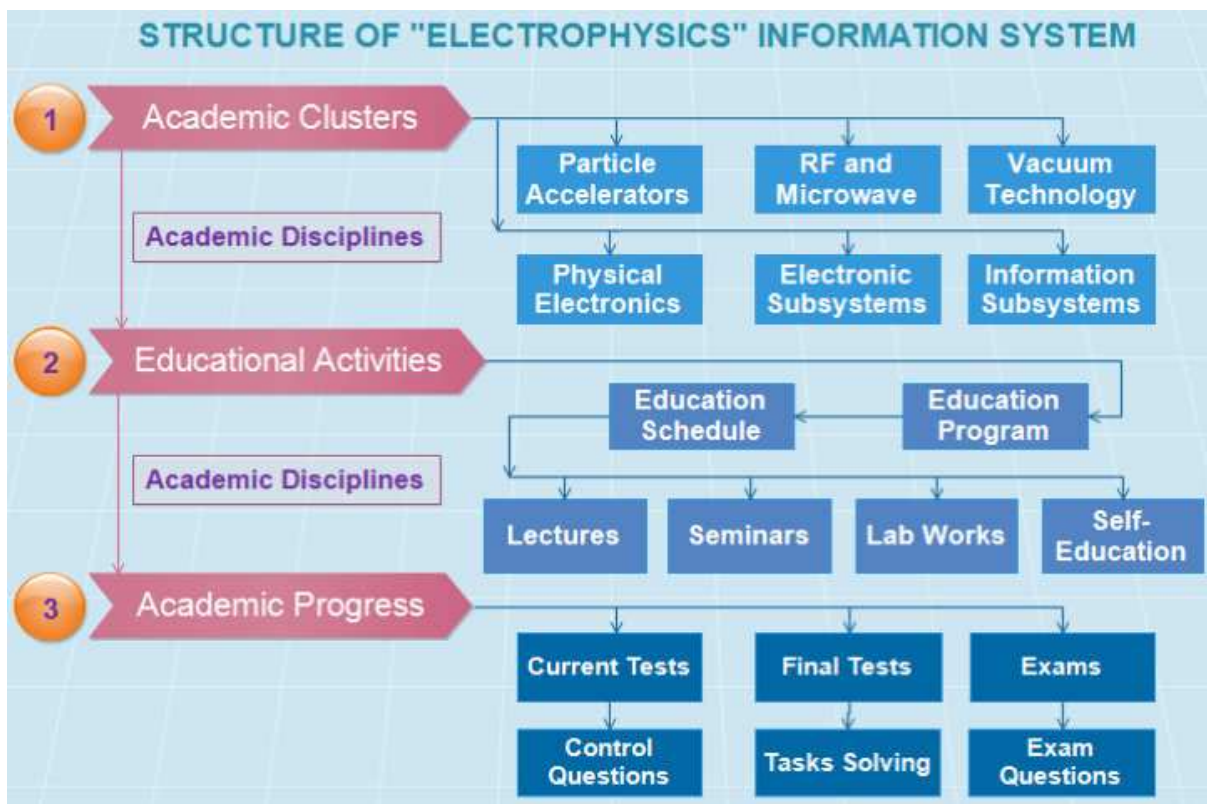}  
\end{center}
\caption{The hierarchical structure of the Information System of INOR ``Electrophysics''.}
\end{figure}

The levels of the hierarchy of the object-oriented aspect are shown in Fig. 2. The first level of the hierarchy of the thematic structure of the information system of INOR ``Electrophysics'' is associated with the distribution of academic disciplines within educational and scientific cycles, defining the main directions of training of specialists in the physics of charged particle beams and accelerator technology area:
\begin{enumerate}
\item  Accelerators of charged particles;
\item  Microwave engineering;
\item  Physical electronics;
\item  Electronic systems of accelerators;
\item  Information systems of accelerators.
\end{enumerate}

The second level is associated with the description of the structure and technologies of the educational process for an individual training module of each training cycle, including training programs, semester schedules, content associated with lectures, practical laboratory classes and self-study. Educational content consists of thematically structured blocks, including e-teaching materials in text and graphical forms (electronic books, textbooks and lecture notes in PDF, DJVU formats, Power Point presentation, etc), as well as video and audio recordings of lectures, practical classes (including webinars) and educational videos.

The third level describes the structure of the tools monitoring of learning progress in the education formats described in the second level. This structure includes various forms of control such as tests, evaluation funds to monitor academic progress, as well as content which contains control questions and tasks both in standard text formats, and in the form of specialized applications for testing knowledge.

\begin{figure}[ht]
\begin{center}
\includegraphics[width=15.cm]{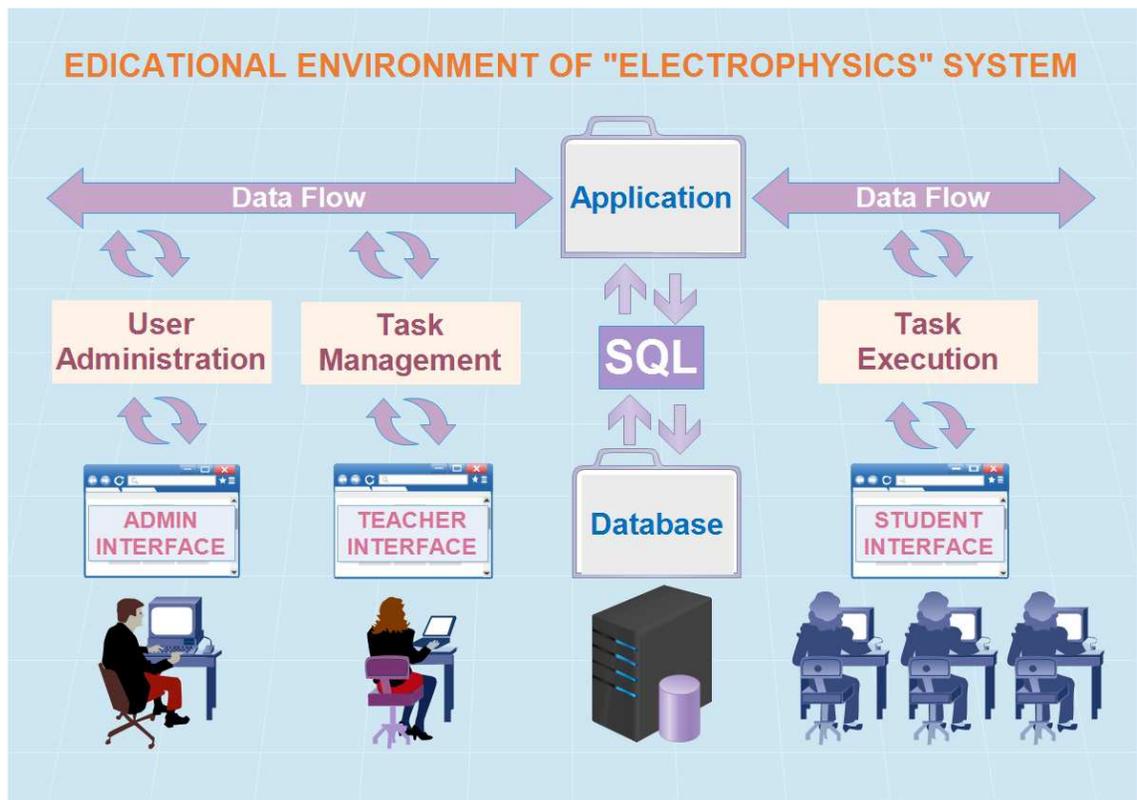}  
\end{center}
\caption{The functional structure of Information System of INOR ``Electrophysics''.}
\end{figure}

Functional or technological aspect is connected with the formalization of algorithms, models and structures of data used in the educational process, and then the inclusion of these formal models in the information environment, a core structure of which is the common database, and associated applications that implement specific functions. Thus, for each functional application, several types of interfaces for different categories of users relevant to their roles and functions in the information system (Fig. 3) are developed.
We can distinguish three main categories (roles) of users in INOR ``Electrophysics'':

\begin{enumerate}
\item  Administrators – monitor relevance of curricula and schedules, as well as reports on current and final certification and the level of access. They manage individual user accounts, form student groups and assign teachers to them, as well as control the level of user access to information resources. 
\item  Teachers - manage the learning process, determine learning objectives in accordance with the curricula, define implementation criteria, form task lists for their student groups, a list of textbooks and references on the teaching content, monitor the assignments implementation and final certification of students.
\item  Students - perform training tasks, having the ability to save the results for further work with them or send them for automated testing or for checks by the tutor, prior to current and, ultimately, final certification.
\end{enumerate}

\section{Virtual Laboratories of Electrophysics}

The core of INOR ``Electrophysics'' are the virtual laboratories based on computer simulation subsystems of charged particles accelerators. The following virtual laboratories have been currently developed and used:
\begin{enumerate}
\item ``The channels of transportation of high-energy particles'';
\item ``Electronic systems of accelerators'';
\item ``Vacuum technique'';
\item ``High-power pulse technology''.
\end{enumerate}

Computer models of subsystems of charged particles accelerators for training students specializing in the field of physics of charged particle beams and accelerator technology have been used at the MEPhI DEF since 1975. This action was forced by circumstances related to the difficulties of implementing full-scale simulation of such subsystems at the teaching laboratories:

\begin{itemize}
\item Laboratory research require unique and expensive equipment;
\item A real subsystem of the electrophysical facilities are of considerable size and high requirements for electrical, electromagnetic, radiation safety;
\item Real experiments take quite a long time and very often are incompatible with the planned schedule of the classes.
\end{itemize}

\begin{figure}[h]
\begin{center}
\includegraphics[width=10.cm]{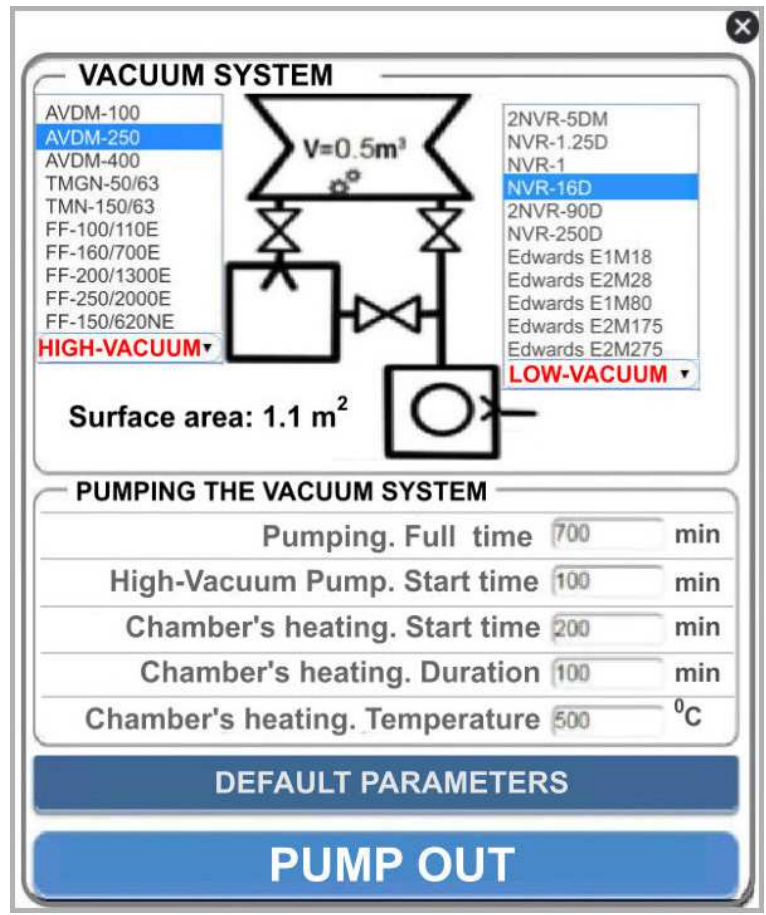}  
\end{center}
\caption{Scheme of vacuum system pumping.}
\end{figure}

In this regard, computer modeling and simulation of elements and subsystems work of electrophysical facilities proved promising, and in some cases is the only possible way to do researches, configure, and design of such devices, partially compensating the lack of real facilities.

Despite the fact that the mathematical models developed for simulation of virtual laboratories, were successfully used in creation of a specialized CAD system for designing subsystems of high-current charged particles accelerators, the concept of virtual lab differs from the concept of CAD. The virtual laboratory is not a tool of direct design, and it is aimed at studying the physical processes in the specified scheme devices under study.

\begin{figure}[h]
\begin{center}
\includegraphics[width=15.cm]{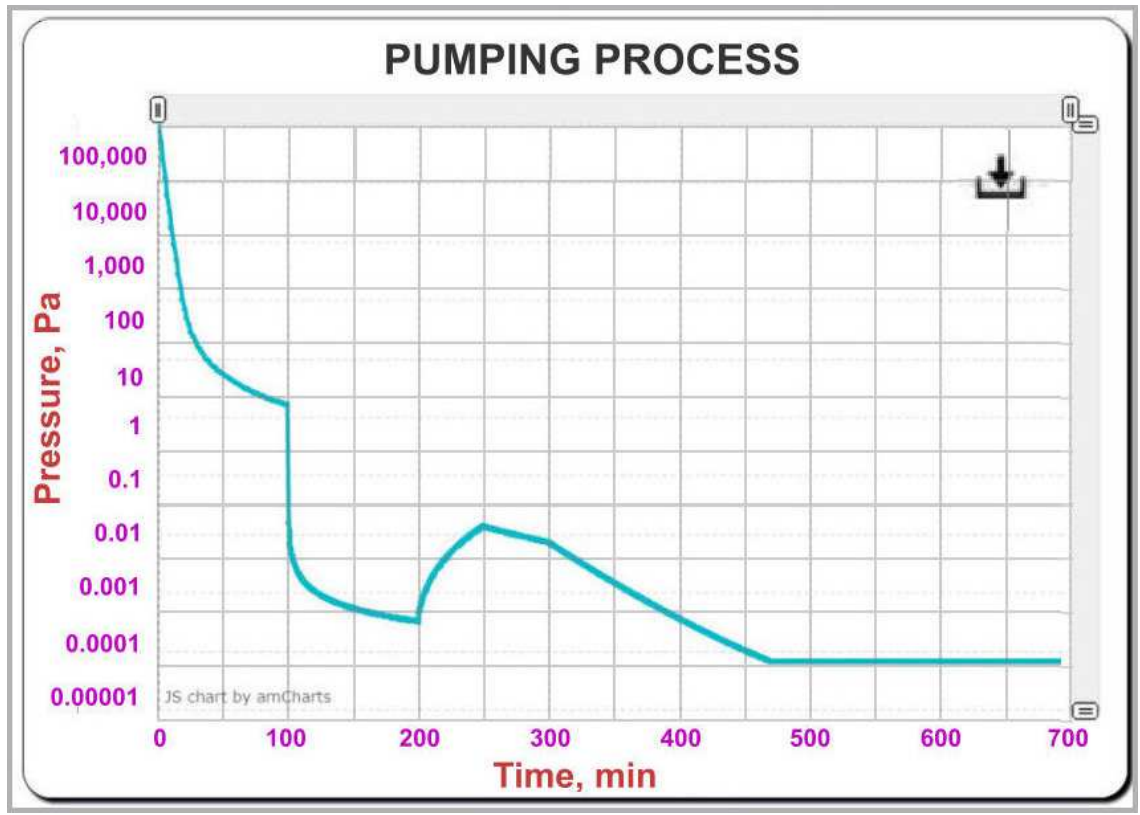}  
\end{center}
\caption{The process of vacuum system pumping.}
\end{figure}

The virtual laboratory allows the assembly of diagrams of the studied devices on the computer monitor with mouse clicks, but the user has to deal with already assembled investigated circuit (Fig. 4), in which it is possible to change only some elements at fixed positions. Otherwise, it works as a CAD software – used to configure the scheme settings, and the result is visualized in graphs of the studied processes (Fig. 5).

Currently, the virtual laboratories have been implemented as cross-platform web application (HTML, CSS, and JavaScript), using AJAX and CGI technologies. In the future, it is planned to implement the concept of Cloud Computing, with allocating computing resources dynamically to each running application.

\section{Conclusion}

The development of INOR ``Electrophysics'' has allowed to conduct a technological audit of the software, and to develop a common conceptual framework for the development and use of information systems in the organization of training of specialists in the ``Physics of charged particle beams and accelerator technology'' area.

High rates of development of modern IT extend the functionalities of the development, increasing the productivity of developers due to new instrumental software. On the other hand, for the same reasons, the developed software quickly becomes outdated, and, actually, its constant modernization is required. The way out of this situation is to create universal cross-platform Web applications that retain relevance when changing the hardware platform.

Distance teaching serves as an excellent additional tool for the empowerment of independent work of students. The introduction of systems of remote training allows improving the manageability and efficiency of the educational process. Performing laboratory workshops at the virtual labs of INOR ``Electrophysics'', students have the opportunity to gain significant practical experience required for designing of subsystems of charged particles accelerators in the future.

Prospects for the development of INOR ``Electrophysics'' on the one hand are connected with the increasing of performance of modern computers that allows improving the accuracy of modeling the physical processes. On the other hand, empowerment of software tools of web development will allow developing universal cross-platform applications for distance teaching, having an adaptive interface. This creates a possibility for remote access to teaching tools using not only desktops with widescreen monitors or laptops, but also mobile devices, contributing to the empowerment of the self-study everywhere (with Internet access) at any time.

\end{document}